\documentclass[global, twocolumn]{svjour}
\usepackage{graphics}
\usepackage{txfonts, multirow, array, booktabs, cite}

\journalname{}

\begin{document}
\title{Non-negligible collisions of alkali atoms with background gas in buffer-gas-free cells coated with paraffin}
\author{Naota Sekiguchi \and Atsushi Hatakeyama}
\institute{Department of Applied Physics,
Tokyo University of Agriculture and Technology,
Koganei, Tokyo 184-8588, Japan\\
\email{hatakeya@cc.tuat.ac.jp}\\
Tel:~+81~(0)~42~388~7554\\
Fax:~+81~(0)~42~388~7554
}

\date{January 13, 2016}

\maketitle

\begin{abstract}
We measured the rate of velocity-changing collisions (VCCs)
between alkali atoms and background gas 
in buffer-gas-free anti-relaxation-coated cells.
The average VCC rate in paraffin-coated rubidium vapor cells 
prepared in this work was $1\times 10^{6}$~s$^{-1}$,
which corresponds to $\sim 1$~mm in the mean free path of rubidium atoms.
This short mean free path indicates that
the background gas is not negligible in the sense that
alkali atoms do not travel freely between the cell walls.
In addition, we found that 
a heating process known as ``ripening''
increases the VCC rate, 
and also confirmed that ripening
improves the anti-relaxation performance of the coatings.
\end{abstract}

\section{Introduction}
\label{sec:intro}
There are two major techniques used to preserve a spin-polarized ground state
of alkali atoms in a glass cell.
The first is 
to fill the cell with an inert buffer gas.
The second is to cover the 
inner glass surfaces with a so-called anti-relaxation coating.
The interaction of the anti-relaxation coating with alkali atoms is much weaker than that of the bare glass surface, 
which causes the alkali atoms to bounce many times without relaxation of spin polarization.
Paraffin, 
one of the most widely used anti-relaxation coating materials,
can bounce alkali atoms up to 10~000 times before wall-induced relaxation\cite{Rob58,Bou66}.
Buffer-gas-free anti-relaxation-coated cells have a number of advantages 
over buffer-gas cells:
the broadening of magnetic resonance lines due to field inhomogeneity can be reduced
because alkali atoms sample the entire volume of the cell, 
and also it is possible to polarize and detect alkali atoms more effectively
because there is no pressure broadening of the optical resonance line.
Buffer-gas-free anti-relaxation-coated cells have therefore been used in many applications, including
atomic clocks\cite{Rob82,Ban12}, 
ultra-sensitive magnetometry\cite{Dup69,Bud00,Bud07}, 
quantum memory\cite{Jul04}, and
studies of light propagation\cite{Bud99}.
In these experiments,
the background gas pressure is thought to be low
enough to ignore collisions between alkali atoms and the background gas.
However, 
the high reactivity of alkali atoms
would lead to the generation of additional gases by chemical reactions.
Are the coated cells really ``buffer-gas-free'' 
in the sense that alkali atoms travel freely between the walls?

In fact, 
outgassing in paraffin coated cells has been concerned\cite{Bou66}
and chemical reactions of alkali atoms
with silicon compound coatings and resulting outgassing
have been reported\cite{Cam87a,Yi08,Atu14,Ste94}.
If the background gas pressure is too high for alkali atoms
to travel freely between the walls,
the influences of the collisions should be taken into account.
For example, 
collisions between cesium atoms and hydrogen molecules, 
which are typical species of the background gas\cite{Cam87a,Yi08,Atu14},
shift the ground state hyperfine transition frequency by 14~Hz/Pa\cite{Ard58}
at 30$^{\circ}$C. 
This shift may be problematic in atomic clock applications,
considering the frequency shift induced by inner surfaces 
of a coated cell is on the order of a hundred Hz\cite{Rob82, Ban12}.
However, background gas pressures in buffer-gas-free cells 
have not been reported 
except for a very high pressure ($\leq 600$~Pa) background gas 
in a microfabricated cell with an octadecyltrichlorosilane coating\cite{Str14}.

In situ pressure measurement without breaking the cell 
makes it possible to track the history of the pressure over the heating process,
sometimes called ``ripening,'' 
which may accelerate chemical reactions.
Ripening is generally performed on coated cells before experiments
to increase alkali vapor density\cite{Bou66,Cam87a,Ale02,Sel13,Gra05},
because the initial alkali vapor density is much lower 
than its saturation vapor density.
It has been reported that 
ripening improves anti-relaxation performance\cite{Cam87a,Gra05},
although there are concerns that ripening may simply reduce the collision rate of atoms
with the cell surface by increasing the background gas pressure\cite{Cam87a}.

Here, 
we report a simple and effective method for measuring the rate of velocity-changing collisions (VCCs)
between alkali atoms and background gas without breaking the cell
when the gas pressure is low enough not to cause pressure broadening.
The background gas pressure and 
mean free path of alkali atoms can be estimated from the measured VCC rate.
This measurement was demonstrated 
in rubidium (Rb) vapor cells with a paraffin coating.
We found that 
the VCC rates were higher than a few $10^{5}$~s$^{-1}$.
The mean free path of Rb atoms and the background gas pressure were
estimated to be a few mm and several Pa, respectively.
In addition, 
by measuring the VCC rates before and after ripening,
we found that ripening increased the background gas pressure.
The spin-relaxation times were also measured before and after ripening, 
and the number of bounces before relaxation was evaluated.
We found that ripening increased the number of bounces.

\section{VCC-rate measurement}
\label{sec:VCC-rate measurement}

\subsection{Method}
\label{sec:method}

\begin{figure}[b]
\centering
\resizebox{0.9\hsize}{!}{%
	\includegraphics{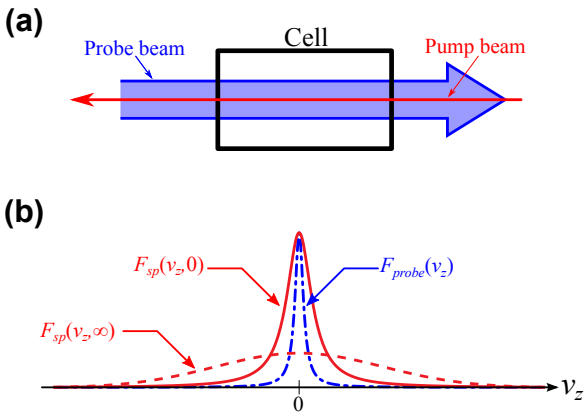}
}
\caption{(Color online) 
Schematic illustration of VCC rate measurement.
(a)
Spin polarization produced by the pump pulse 
is detected with the probe beam, 
which is counterpropagating and collinear to the pump beam.
(b)
The probe beam detects the spin polarization 
of atoms within the velocity distribution $F_{probe}(v_{z})$.
The initial velocity distribution $F_{sp}(v_{z},0)$ of spin polarization, 
which can be broader that $F_{probe}(v_{z})$ because of power broadening,
is redistributed toward $F_{sp}(v_{z},\infty )$.
}
\label{fig:concept}
\end{figure}

Our method to measure the VCC rate $\gamma$
is based on velocity-selective optical pumping 
and VCC-induced redistribution of the velocity of spin-polarized alkali atoms.
A resonant, circularly polarized pump laser pulse
selectively polarizes alkali atoms 
whose velocity components $v_{z}$ along the laser propagation direction (the $z$ direction)
lie within a narrow velocity distribution centered at $v_{z}=0$.
The pump pulse irradiates alkali atoms for only a short duration.
Spin polarization is detected with a counterpropagating collinear probe laser beam
(see Fig.~\ref{fig:concept}a).
The diameter of the probe beam is much larger than that of the pump beam.
A similar optical configuration has been adopted to measure diffusion coefficients\cite{Par14}.

The signal $S(t)$
measured with the probe beam indicates the polarization of atoms with $v_{z}\sim 0$ in the probe beam region and
decreases with time $t$
due to two different mechanisms.
First, the ballistic escape of the polarized atoms from the probe beam region
decreases $S(t)$.
Second, the VCCs between alkali atoms and the background gas 
causes a decrease in $S(t)$,
as the VCCs redistribute the atomic velocity of the polarized atoms toward
the Maxwell-Boltzmann distribution,
resulting in fewer atoms within a probed velocity range
(Fig.~\ref{fig:concept}b).
The signal $S(t)$ is thus described by
\begin{equation}
S(t) \propto S_{esc}(t)\times S_{c}(t),
\label{eq: signal_evolution}
\end{equation}
where the contributions of the ballistic escape and the VCCs are denoted as
$S_{esc}(t)$ and $S_{c}(t)$, respectively.
Detailed discussions about $S_{esc}(t)$ and $S_{c}(t)$ will be provided in Sec.~\ref{sec:analysis}.

\subsection{Experimental setup}
\label{sec:VCC setup}
Measurement of the VCC rate was carried out 
for paraffin-coated Rb vapor cells at room temperature,
the fabrication procedure of which will be described in Sec.~\ref{sec:experiment}.
A longitudinal magnetic field of 40~$\muup$T was applied to the vapor cell with Helmholtz coils 
in a permalloy magnetic shield.
The frequencies of the pump and the probe beams 
were stabilized to the $F=3\to F'=4$ transition frequency of the $^{85}$Rb $D2$ line.
The profiles of the beams were measured with a CMOS camera 
and fitted by Gaussian functions.
The diameters of the probe and the pump beams 
at which their intensities were reduced by $1/e^{2}$
compared to the beam center intensity
were 0.38~cm and 0.053~cm, respectively. 
The probe beam power was 10~$\muup$W, and 
the pump beam power was 0.3~mW.

The pump pulse has a finite duration $\tau$, 
and $\tau$ must be much longer than the lifetime of the Rb $D2$ line
to polarize Rb atoms.
At the same time, $\tau$ should not be much longer
than the typical time constants of the atoms' ballistic escape 
and the decrease in $S(t)$ due to the VCCs.
In this experiment, the pump beam irradiated the cell for 1~$\muup$s,
and was then shut off with an acousto-optic modulator at $t=0$.
Measurements were performed for the left- and right-circularly polarized probe beams, 
and the signal $S(t)$ 
was derived from the difference between the transmitted powers 
for the two polarizations.

\subsection{Analysis}
\label{sec:analysis}
\begin{figure}[htbp]
\centering
\resizebox{1.0\hsize}{!}{%
	\includegraphics{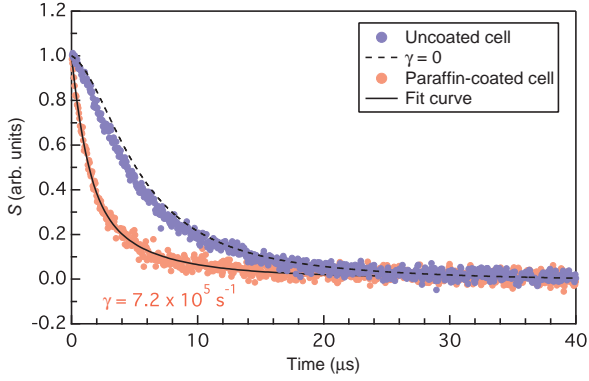}
}
\caption{(Color online)
Signals $S$ of the VCC rate measurement 
for an uncoated cell shown by blue points and 
a paraffin-coated cell (\#2, post-ripening) shown by red points.
Both data are normalized at 0~$\muup$s.
The dashed line shows a simulation of the $S$ without VCCs.
The solid line shows a fit curve to the data for the coated cell. 
}
\label{fig:uncoatedcell}
\end{figure}

Experimental data for an uncoated cell and a paraffin-coated cell
are shown in Fig.~\ref{fig:uncoatedcell}
by blue and red points, respectively.
It is clear that the decrease in signal $S(t)$ for the paraffin-coated cell was more rapid 
than that for the uncoated cell.
The paraffin-coated cell, therefore, contained more background gas than the uncoated cell.
To derive the VCC rate,
we simulated the signal $S(t)$ with the following model.

First, we considered the contribution $S_{esc}(t)$ of the ballistic escape to the signal $S(t)$.
The polarized atoms' velocity component $v_{r}$ perpendicular to the beams 
obeys the Maxwell-Boltzmann distribution.
For simplicity, we assumed that
the pump beam diameter was infinitely small 
and atoms were polarized at the center of the probe beam at $t=0$.
The probability $G(v_{r})dv_{r}$ 
that a polarized atom has a velocity between $v_{r}$ and $v_{r}+dv_{r}$ was yielded as 
\begin{equation}
G(v_{r}) dv_{r} = \frac{2v_{r}}{v_{D}^{2}} \exp \left \{-\left (\frac{v_{r}}{v_{D}}\right )^{2} \right \} dv_{r},
\label{eq:Boltzmanndis}
\end{equation} 
where the most probable speed $v_{D}$ is given by $v_{D} = \sqrt{2k_{B}T/m}$
with the Boltzmann constant $k_{B}$, the absolute temperature $T$, and the atomic mass $m$.
The polarized atoms move at a constant velocity during the ballistic escape.
Hence, the probability $\rho (r,t)dr$ 
that a polarized atom is at a distance from the center between $r$ and $r + dr$ at time $t$
was derived from Eq.~(\ref{eq:Boltzmanndis}) as
\begin{equation}
\rho(r,t)dr = \frac{2r}{(v_{D}t)^{2}}\exp \left\{ -\left( \frac{r}{v_{D}t}\right)^{2} \right\} dr.
\label{eq:probability}
\end{equation}

As the optical depths of the vapor cells were much smaller than unity at room temperature, 
the signal is proportional to the product
of $\rho(r,t)dr$ and the intensity 
of the probe beam at the position of a polarized atom:
\begin{equation}
S_{esc}(t) \propto \int_{0}^{\infty} I_{probe}(r)\times \rho(r,t) dr.
\label{eq:signal_esc}
\end{equation}
Here, the beam profile $I_{probe}(r)$ of the probe beam was described by a Gaussian function 
\begin{equation}
I_{probe}(r) \propto \exp \left \{ - \left ( \frac{\sqrt{2}r}{\phi_{probe}/2} \right )^{2} \right \},
\label{eq:probe_profile}
\end{equation}
where the probe beam diameter 
$\phi_{probe}=0.38$~cm,
as mentioned above.
By substituting Eqs.~(\ref{eq:probability}),~(\ref{eq:probe_profile}) into Eq.~(\ref{eq:signal_esc}),
$S_{esc}(t)$ was obtained as
\begin{equation}
S_{esc}(t) \propto \frac{1}{\phi_{probe}^{2}+8v_{D}^{2}t^{2}}.
\label{eq:escape}
\end{equation}

Next, we considered the contribution $S_{c}(t)$ of the VCCs to the signal $S(t)$.
The polarization of the alkali atoms has a velocity distribution $F_{sp}(v_{z})$.
The velocity distribution $F_{sp}(v_{z})$ are redistributed toward the Maxwell-Boltzmann distribution
due to the VCCs,
while the polarization is preserved.
The velocity distribution $F_{sp}(v_{z})$ of the spin polarization, thus, depends on $t$ 
and is described as $F_{sp}(v_{z},t)$.
The pump pulse beam produces the initial Lorentzian distribution $F_{sp}(v_{z},t=0)$,
which corresponds to the absorption profile for the pump laser.
In the steady state, $F_{sp}(v_{z},t=\infty)$ should be proportional to 
the Maxwell-Boltzmann distribution.
The probe beam detects the polarization
within the velocity distribution $F_{probe}(v_{z})$, 
which is proportional to the absorption cross-section for the probe beam.
The signal detected with the probe beam is proportional to the integral
of the product of $F_{sp}(v_{z},t)$ and $F_{probe}(v_{z})$ over the velocity $v_{z}$,
\begin{equation}
S_{c}(t)\propto \int_{-\infty}^{\infty} F_{sp}(v_{z},t) \times F_{probe}(v_{z})dv_{z}.
\label{eq:VCCs}
\end{equation} 
We assumed the time variation of $F_{sp}(v_{z},t)$ as
\begin{equation}
\frac{\partial F_{sp}(v_{z},t)}{\partial t} =
-\gamma \left[
F_{sp}(v_{z},t) - F_{sp}(v_{z},\infty)
\right],
\label{eq:redistribution}
\end{equation}
where the steady-state velocity distribution $F_{sp}(v_{z},\infty)$ was given by
\begin{equation}
F_{sp}(v_{z},\infty)=\int_{-\infty}^{\infty} F_{sp}(v_{z}',0)dv_{z}'\times \frac{1}{v_{D}\sqrt{\pi}} 
\exp \left\{ -\left( \frac{v_{z}}{v_{D}}\right) ^{2} \right\}.
\label{eq:steady state}
\end{equation}
We then calculated $S_{c}(t)$ 
from Eqs.~(\ref{eq:VCCs}),~(\ref{eq:redistribution}),~and~(\ref{eq:steady state}).

As the pump pulse has a finite duration $\tau=1$~$\muup$s,
the experimental data were compared with
$\int_{-\tau}^{0}S_{esc}(t-t')\times S_{c}(t-t')dt'$.
The curve calculated with the VCC rate $\gamma=0$ (the dashed line in Fig.~\ref{fig:uncoatedcell})
in the absence of VCCs
is in reasonable agreement with the experimental data for the uncoated cell.
The experimental data for the coated cell were well fitted by the curve
with $\gamma=7.2\times 10^{5}$~s$^{-1}$ (the solid line in Fig.~\ref{fig:uncoatedcell}).
In this work, 
we consider that
dominant uncertainties in the VCC rate $\gamma$ come from 
our adopted model or uncertainties in the experimental conditions,
while the fitting errors are on the order of 1\%.

\subsection{Estimation of background gas pressure and mean free path of alkali atoms}
\label{sec:estimation}
The VCC rate $\gamma$ is given by $\gamma = n \sigma_{c} \bar{v}$.
Here, $n$ is the number density of background gas, 
$\sigma_{c}$ is the cross-section of VCCs, 
and $\bar{v}$ is the mean relative velocity.
The mean free path $\lambda$ of Rb atoms is then $\lambda = 1/(n\sigma_{c}) = \bar{v}/\gamma$.
Although the background gas species were unknown in this work, 
we assumed $\bar{v}=1\times 10^{5}$~cm/s
as most components of the background gas 
in some silicon compound-coated cells are light\cite{Cam87a,Yi08,Atu14}.
From the ideal gas law, the
pressure $p$ of the background gas was estimated by 
$p=k_{B}T/(\lambda \sigma_{c})$, 
where $\sigma_{c}$ was assumed to be $\sigma_{c}=1\times 10^{-14}$~cm$^{2}$
by reference to Ref.~\citen{Ami83} 
regarding Na-Ne collisions.
The mean free path and background gas pressure 
for the paraffin-coated cell shown in Fig.~\ref{fig:uncoatedcell}
were estimated to be 1.4~mm and 3.0~Pa, respectively.

\section{Cell preparation and spin-relaxation measurement}
\label{sec:experiment}

\begin{table}[htbp]
\centering
\caption{%
Dimensions and slow spin-relaxation time $\tau_{s}$ of cylindrical paraffin-coated cells.
The coated cells are numbered from \#1 to \#7.}
\newcolumntype{C}{>{\centering\arraybackslash}p{5mm}}
\newcolumntype{Y}{>{\centering\arraybackslash}p{11mm}}
\newcolumntype{R}{>{\centering\arraybackslash}p{18mm}}
\begin{tabular}{CYYRR}\toprule
\multirow{2}{*}{Cell}&
Diameter&
Length&
\multicolumn{2}{c}{%
slow spin-relaxation time $\tau_{s}$ (s)%
}\\\cmidrule{4-5}
&
(mm)&(mm)&
Pre-ripening&Post-ripening\\\midrule
\#1& 17.0 & 24.4& 0.15&0.18\\
\#2& 17.0 & 27.2& 0.046& 0.082\\
\#3& 17.0 & 27.4& 0.037& 0.15\\
\#4& 17.0 & 25.0& 0.040& 0.19\\
\#5& 17.0 & 24.3& 0.17& 0.22\\
\#6& 17.0 & 23.7& 0.22& 0.24\\
\#7& 17.0 & 24.3& 0.081&0.13\\
\bottomrule
\end{tabular}
\label{tb:cells}
\end{table}

The above evaluations were performed for seven paraffin-coated cells
made of Pyrex glass.
The dimensions of the coated cells are shown in Table~\ref{tb:cells}.
Each cell had a stem with a constricted capillary with a diameter of $\sim 1$~mm
to be connected to a vacuum system.
The coating procedure was similar to those reported previously\cite{Ale02,Sel13}.
Coating material was fractional distilled between 220$^{\circ}$C and 240$^{\circ}$C 
from commercial paraffin wax 
Sasolwax\textsuperscript{\textregistered} H1 produced by Sasol Wax.
The cell was evacuated below $10^{-4}$~Pa with a turbomolecular pump and 
baked at 420$^{\circ}$C for 4 hours.
After deposition of paraffin at 380$^{\circ}$C or 400$^{\circ}$C, 
the cell was re-evacuated below $10^{-4}$~Pa 
and then sealed with Rb metal (in natural isotopic abundance) at the stem.
In this work, initial Rb vapor density 
was sufficiently high to carry out the VCC rate and spin-relaxation time measurements.
The measurements were also performed after ripening,
during which the cell was heated at 80$^{\circ}$C for 10~hours
while the stem was kept at 75$^{\circ}$C.

\begin{figure}[b]
\centering
\resizebox{1.03\hsize}{!}{%
	\includegraphics{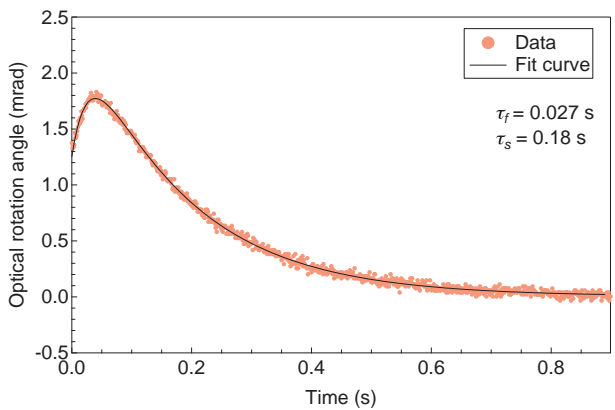}
}
\caption{(Color online)
Spin relaxation time measurement in a paraffin-coated cell
(\# 1, post-ripening).
The experimental data were fitted by a double exponential function.
}
\label{fig:Cell1_T1}
\end{figure}

The longitudinal spin-relaxation times of the fabricated cells were measured 
in a manner similar to that described previously\cite{Gra05}.
The cell was placed in the magnetic shield and immersed in a longitudinal magnetic field
of $40$~$\muup$T.
A circularly polarized laser light (pump light) produced spin polarization of Rb atoms 
and was shut off at time $t=0$.
A linearly polarized laser light (probe light) and a balanced polarimeter detected 
the optical rotation induced by spin polarization.
The power of the probe light was 5~$\muup$W, 
and the probe beam profile was the same as that used in VCC-rate measurement.
Figure~\ref{fig:Cell1_T1} shows the time dependence of the optical rotation angle
in a paraffin-coated cell (\#~1, post-ripening).
The angle of the polarization plane of the probe light 
for the unpolarized atoms was adjusted to 0.0~mrad.
The data were fitted by a double exponential function and 
indicated two relaxation components, 
a fast relaxation time $\tau_{f}$ and a slow relaxation time $\tau_{s}$, 
with opposite signs 
as reported previously\cite{Gra05}.
We focused on the slow relaxation time $\tau_{s}$,
which is shown in Table~\ref{tb:cells}.

\begin{figure}[b]
\centering
\resizebox{0.95\hsize}{!}{%
	\includegraphics{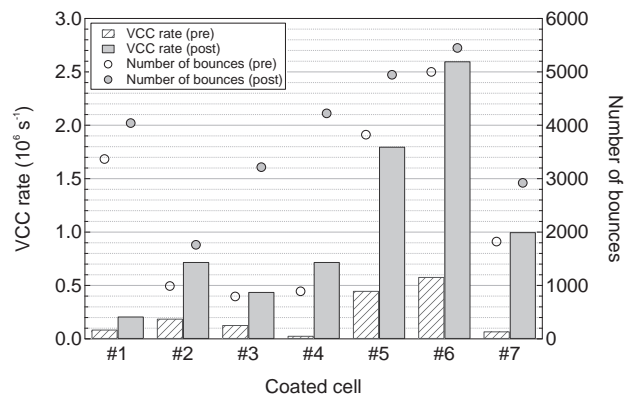}
}
\caption{VCC rates (left axis) and number of bounces (right axis) 
for pre- and post-ripening coated cells.
Hatched bars show VCC rates for pre-ripening cells, 
and gray bars show those for post-ripening cells.
The numbers of bounces for pre- and post-ripening cells are shown by
open circles and gray circles, respectively.
}
\label{fig:RG}
\end{figure}

\section{Results and Discussion}
\label{sec:result}
VCC rates in all prepared coated cells are shown in Fig.~\ref{fig:RG}.
Although there was variation in VCC rate among the coated cells, 
VCC rates in the post-ripening coated cells
were increased to higher than a few $10^{5}$~s$^{-1}$.
The corresponding mean free paths
were shorter than a few~mm,
which was much smaller than the dimensions of a typical cell.
It is clear that 
the ripening process made the background gas pressure higher in all coated cells.

To test the generality of our results,
we also measured the VCC rate in a pwMB-coated cell fabricated in a different laboratory.
The coating material pwMB was mentioned in Refs.~\citen{Sel10,Sel13,Hib13}.
The pwMB-coated cell was spherical (47~mm in diameter)
and filled with enriched $^{85}$Rb.
The VCC rate in the pwMB-coated cell was measured as $3.8\times 10^{5}$~s$^{-1}$.
The mean free path and the background gas pressure were estimated to be 2.6~mm and 1.6~Pa, respectively.
It is therefore very likely that 
paraffin-coated cells generally have relatively high background gas pressures.

With the VCC rates in mind,
we evaluated the number of bounces $N$ before depolarization
under the assumption that Rb atoms moved diffusively
as described previously\cite{Sel08}.
In the evaluation, 
the diffusion coefficient $D$ had two variations. 
One was estimated from the classical diffusion coefficient 
$D=\lambda v_{m}/3$, where the mean speed $v_{m}$ of Rb atoms is given by
$v_{m} = 2v_{D}/\sqrt{\pi}$.
The other was estimated from the diffusion coefficient for Rb atoms in hydrogen molecules\cite{McN62}
and the estimated background gas pressure.
In addition, 
the number of bounces in the absence of VCCs was also calculated by
\begin{equation}
N = \frac{v_{m}A}{4V}\times \tau_{s}.
\end{equation}
Here, $A$ is the inner surface area and $V$ is the volume of the cell.
The numbers evaluated in these three ways were almost the same, 
as expected for the background gas pressure of a few Pa,
and are shown in Fig.~\ref{fig:RG}.
It is clear that
the number of bounces, i.e., the anti-relaxation performance of the coating,
was improved by ripening.

\section{Conclusions}
\label{sec:conclusion}
We measured the rates of VCCs 
between Rb atoms and background gas in seven buffer-gas-free, paraffin-coated cells.
The results indicated 
an average VCC rate of $1\times 10^{6}$~s$^{-1}$.
This VCC rate corresponds to 
the mean free path of 1~mm and the background gas pressure of 4~Pa.
This short mean free path compared to the cell dimensions means that
Rb atoms in the coated cells do not travel freely between the cell walls.
We also found that
the so-called ``ripening'' process increased the background gas pressure.
We confirmed improvement of the anti-relaxation performance of the paraffin coatings 
by ripening.
The relatively high VCC rates indicated that precise experiments using coated cells such as clock applications 
require attention to the effects induced by collisions with background gas. 
Models to describe atomic dynamics in coated cells, 
such as discussed in Ref.~\citen{Kle11} to reproduce electromagnetically induced transparency spectra, 
may require modification to the assumption of the ballistic transport of atoms in the cells.
In particular, 
the background gas pressure should be checked when coated cells are heated.

\begin{acknowledgement}
	We would like to thank Mikhail~V.~Balabas 
	for fabrication of the pwMB-coated cell and for his helpful comments, 
	and Antoine Weis for his important advice.
	This work was supported by a Grant-in-Aid for Scientific Research (No.~23244082)
	from the Japan Society for the Promotion of Science (JSPS).
\end{acknowledgement}

\end{document}